\documentclass[graybox]{svmult}

\usepackage{mathptmx}       
\usepackage{helvet}         
\usepackage{courier}        
\usepackage{type1cm}        
\usepackage{makeidx}  
\usepackage{graphicx}        
\usepackage{multicol}        
\usepackage[bottom]{footmisc}

\makeindex             

\begin{document}

\title*{Spreading, Nonergodicity, and Selftrapping: a puzzle of interacting disordered lattice waves  }
\author{Sergej Flach}
\institute{Sergej Flach \at 
Center for Theoretical Physics of Complex Systems, Institute for Basic Science, Daejeon, Korea, \email{sflach@ibs.re.kr} \\
New Zealand Institute for Advanced Study, Centre for Theoretical Chemistry and Physics, Massey University, 0745 Auckland New Zealand. \email{s.flach@massey.ac.nz}
}
%
%
\maketitle

\abstract{Localization of waves by disorder is a fundamental physical problem encompassing a diverse spectrum of theoretical, experimental and numerical studies in the context of metal-insulator transitions, the quantum Hall effect, light propagation in photonic crystals, and dynamics of ultra-cold atoms in optical arrays, to name just a few examples. Large intensity light can induce nonlinear response, ultracold atomic gases can be tuned into
an interacting regime, which leads again to nonlinear wave equations on a mean field level.
The interplay between \index{disorder} disorder and \index{nonlinearity} nonlinearity, their localizing and delocalizing effects is currently an intriguing and challenging issue in the field of lattice waves. In particular it leads to the prediction and observation of two different regimes of destruction of Anderson localization - asymptotic weak chaos, and intermediate strong chaos, separated by a crossover condition on 
densities. 
On the other side approximate full quantum interacting many body treatments were recently used to predict and obtain
a novel many body localization transition, and two distinct phases - a localization phase, and a delocalization phase, both again separated
by some typical density scale.
We will discuss selftrapping, nonergodicity and nonGibbsean phases which are typical for such discrete models with particle number conservation and their relation to the above crossover and transition physics. We will also discuss potential connections to quantum many body theories.
}


\section{Introduction}
\label{sec1}

\index{Anderson localization}
In this contribution we will discuss the regimes of wave packet spreading in nonlinear disordered lattice systems, and its
relation to quantum many body localization (MBL).
We will consider cases when the corresponding linear (single particle) wave equations 
show Anderson localization, and the localization length \index{localization length} is bounded from above by a finite value.

There are several reasons to analyze such situations. Wave propagation in spatially disordered
media has been of practical interest since the early times of studies of conductivity in solids. In particular,
it became of much practical interest for the conductance properties of electrons in semiconductor devices more
than half a century ago. It was probably these issues which motivated P. W. Anderson to perform his
groundbreaking lattice wave studies on what is now called Anderson localization \cite{PWA58}. With evolving modern technology, wave propagation
became of importance also in photonic and acoustic devices in structured materials \cite{Exp,Exp2}. Finally, recent advances in the control of
ultracold atoms in optical potentials made it possible to observe Anderson localization there as well \cite{BECEXP}.

In many if not all cases wave-wave interactions can be of importance, or can even be controlled experimentally.
Short range interactions hold for s-wave scattering of atoms. When many quantum particles interact weakly,
mean field approximations often lead to effective nonlinear wave equations. Electron-electron interactions in solids and mesoscopic devices are also interesting candidates
with the twist of a new statistics of fermions. 
Nonlinear wave equations
in disordered media are of practical importance also because of  high intensity light beams propagating through structured optical devices
induce a nonlinear response of the medium and subsequent nonlinear contributions to the light wave equations.
While electronic excitations often suffer from dephasing due to interactions with other degrees of freedom (e.g. phonons),
the level of phase coherence can be controlled in a much better way for ultracold atomic gases and light.

There is a fundamental mathematical interest in the understanding, how Anderson localization is modified 
in the presence of quantum many body interactions, or classical nonlinear terms in the wave equations. All of the above motivates the choice of corresponding
linear (single particle) wave equations with finite upper bounds on the localization length. Then, the corresponding
noninteracting quantum systems, as well as linear classical waves,
admit no transport. Analyzing transport properties of nonlinear, respectively interacting quantum, disordered wave equations allows to 
observe and characterize the influence of wave-wave interactions on Anderson localization in a straightforward
way. 

The chapter is structured in the following way. We will introduce the classical model, discuss its statistical properties, and in particular the nonGibbsean phase, and some of its
generalizations.
We will then come to wave packet spreading, a self-trapping theorem, and to results of destruction of Anderson localization. 
Finally, we will discuss the relation of these results to many body localization for the corresponding quantum many body problems.

\section{The model}
\label{sec2}

The Gross-Pitaevskii equation is describing Bose-Einstein condensates (BEC) of interacting ultracold atoms in certain mean-field approximations. It is also known
as the nonlinear Schr\"odinger equation which is integrable in 1+1 dimensions, and has many further applications e.g. in nonlinear optics.
Its discretized version - the discrete nonlinear Schr\"odinger equation (DNLS) or discrete Gross-Pitaevskii equation (DGP) - is typically nonintegrable, and is realized with a BEC loaded onto optical lattices \cite{oberthaler}. Similar the discrete nonlinear Schr\"odinger equation
(DNLS) is realized for various one- and two-dimensional networks of interacting optical waveguides \cite{yskgpa03}. 

Additional disorder, either due to natural inhomogeneities, or intentionally 
implanted, finally leads to the disordered discrete Gross-Pitaevsky (dDGP) or disordered discrete nonlinear Schr\"odinger equation
(dDNLS) equation with Hamiltonian
\begin{equation}
\mathcal{H}= \sum_{l} \epsilon_{l} 
|\psi_{l}|^2+\frac{\nu}{2} |\psi_{l}|^{4}
- (\psi_{l+1}\psi_l^*  +\psi_{l+1}^* \psi_l)
\label{RDNLS}
\end{equation}
with complex variables $\psi_{l}$, lattice site indices $l$ and nonlinearity \index{nonlinearity}
strength $\nu \geq 0$.   The (typically uncorrelated) random on-site energies $\epsilon_{l}$ with zero average $\bar{\epsilon}= \lim_{N\rightarrow \infty} N^{-1} (\sum_{l=1}^{l=N} \epsilon_l) = 0$ and finite variance $\sigma_{\epsilon} = \lim_{N\rightarrow \infty} N^{-1} (\sum_{l=1}^{l=N} \epsilon_l^2)$
are distributed with some probability density distribution (PDF) $\mathcal{P}_{\epsilon}$. Here we will use the box distribution
$\mathcal{P}_{\epsilon}(x) = \frac{1}{W}$ for $|x| \leq \frac{W}{2}$ and $\mathcal{P}_{\epsilon}=0$ otherwise.

The equations of motion are generated by $\dot{\psi}_{l} = \partial
\mathcal{H}_{D}/ \partial (i \psi^{\star}_{l})$:
\begin{equation}
i\dot{\psi_{l}}= \epsilon_{l} \psi_{l}
+\nu |\psi_{l}|^{2}\psi_{l}
-\psi_{l+1} - \psi_{l-1}\;.
\label{RDNLS-EOM}
\end{equation}
Eqs.~(\ref{RDNLS-EOM}) conserve the energy (\ref{RDNLS}) and the norm $\mathcal{A}
= \sum_{l}|\psi_l|^2$.  Note that varying the norm of an
initial wave packet is strictly equivalent to varying $\nu$.
Note also that the transformation $\psi_l \rightarrow (-1)^l \psi^*_l$, $\nu \rightarrow -\nu$, $\epsilon_l \rightarrow -\epsilon_l$ leaves the
equations of motion invariant. Therefore the sign of the nonlinear coefficient $\nu$ can be fixed without loss of generality to be positive.

\section{Gibbsean and nonGibbsean regimes}
\label{sec3}

The existence of the second - in addition to the energy - conserved quantity $\mathcal{A} \geq 0$, combined with the discreteness-induced bounded kinetic energy
part in (\ref{RDNLS}), has a profound impact on the statistical properties of
the DGP system. At variance to its space-continuous counterparts, the disorder-free translationally invariant lattice model with $\epsilon_l=0$ shows a non-Gibbsean phase
\cite{rckgj00,jr04}. It is separated from the Gibbsean phase by states of infinite temperature. While average densities in the non-Gibbsean phase can be formally described by
Gibbs distributions with negative temperature, in truth the dynamics shows a separation of the complex field $\psi_l$ into a two-component one - a first component of high density
localized spots and a second component of delocalized wave excitations with infinite temperature \cite{rckgj00,jr04,br07,br08,br09}. The high density localized spots are conceptually
very similar to selftrapping and discrete breathers \cite{DB}.

We will briefly compute the effect of nonvanishing disorder $\epsilon_l \neq 0$ on the infinite temperature separation line between the Gibbsean and the non-Gibbsean phases. While the final result was listed in Ref. \cite{dmb14} in an appendix, no details were provided.
We will use the notations of the work by Johansson and Rasmussen \cite{jr04}, to which we refer the reader for further details.
We will also generalize to other types of potentials.

We define the local norm per site as $A_l \geq 0$ and the local phase $0 \leq \phi_l \leq 2\pi$ such that
\begin{equation}
\psi_l = \sqrt{A_l}  {\rm e}^{i \phi _l}\;.
\label{trafo}
\end{equation}
The Hamiltonian transforms into
\begin{equation}
\mathcal{H} = \sum_l \epsilon_l A_l - 2\sqrt{A_lA_{l+1}} \cos (\phi_l - \phi_{l+1} )+ \frac{\nu}{2} A_l^2 \;,
\label{Hamtrafo}
\end{equation}
and the norm simply becomes
\begin{equation}
\mathcal{A} = \sum_l A_l \;.
\label{Normtrafo}
\end{equation}
Assuming a large system of $N$ sites the corresponding average densities become 
\begin{equation}
h = \mathcal{H}/N\;,\; a = \mathcal{A} / N \;.
\label{densities}
\end{equation}
From here on we will always implicitly consider the thermodynamic limit $N \rightarrow \infty$.
We will as well express final results in terms of densities scaled with the nonlinearity parameter $\nu$ \cite{dmb14}:
\begin{equation}
y=\nu h \;,\; x=\nu a \;.
\label{scaleddensities}
\end{equation}
A number of questions can be posed. First, can any pair of realizable densities $\{x,y\}$ be obtained by assuming a Gibbs distribution 
\begin{equation}
\rho_G = \frac{1}{\mathcal{Z}} {\rm e} ^{-\beta(\mathcal{H} + \mu \mathcal{A})}
\label{gibbs}
\end{equation} 
where $\mathcal{Z}$ is the partition function, $\beta$ the inverse temperature, and $\mu$ the chemical potential ? 
And if not, what is the dynamics in the corresponding nonGibbsean phase?

The lowest energy state $E_{min}$ is evidently given by $\phi_l = const$ : 
\begin{equation}
E_{min} =  \sum_l (\epsilon_l - 2) A_l + \frac{\nu}{2} A_l^2 \;,
\label{minenergy}
\end{equation}
In the absence of any potential $\epsilon_l=0$ the lowest energy state is homogeneous: $A_l=\mathcal{A}/n = a$. With the notation of the inverse temperature $\beta$  we conclude that the
zero temperature limit of the scaled energy density at a given value of the scaled norm density is given by  
\begin{equation}
y_{\beta\rightarrow \infty} = -2x+x^2/2\;.
\label{betainfinity}
\end{equation}
For a nonzero potential with finite variance the lowest energy density limit will be lowered by a finite value.
With the distribution $\mathcal{P}_{\epsilon}(x)=\frac{1}{W}$ for $|x| \leq \frac{W}{2}$ we arrive at the upper and lower bound s
\begin{equation}
 -(2+W)x+x^2/2 \leq y_{\beta \rightarrow \infty} \leq -2x + x^2/2\;.
\label{betainfinitydisorder}
\end{equation}
At the same time the upper limit for the total energy of a finite system with $N$ sites is obtained by concentrating all the norm on one
lattice site which yields $E_{max} = \nu \mathcal{A}^2/2$ and an energy density $y_{max} = N x^2/2$. In the thermodynamic limit $N\rightarrow \infty$ the upper limit for
the energy density is diverging. We conclude that at any given norm density $x$ the energy density is bounded from below by the finite minimum value $y_{\beta \rightarrow \infty}$, but
is not bounded from above and can take arbitrary large values.

The partition function 
\begin{equation}
{\mathcal Z}=\int_0^\infty \int_0 ^{2\pi} \prod_m d\phi_m dA_m
\exp[-\beta({\mathcal H}+\mu \mathcal{A})]
\label{partitionfunction}
\end{equation}
depends on all amplitudes and phases.
Integration over the phase variables $\phi_m$  reduces the symmetrized
partition function to
\begin{equation}
{\mathcal Z}=(2\pi)^N\int_0^\infty \prod_m dA_m
I_0(2\beta\sqrt{A_mA_{m+1}})
{\rm e}^{-\beta
(\mathcal{H}_0 + \mu \mathcal{A})}\
\label{reducedpf}
\end{equation}
with the reduced Hamiltonian 
\begin{equation}
\mathcal{H}_0= \sum_l \epsilon_l A_l + \frac{1}{2} A_l^2 
\label{H0}
\end{equation}
depending only on the amplitudes, and with $I_0$ being the Bessel function of 0th order.

The strategy of finding a nonGibbsean phase is simply to find the line of infinite temperature $\beta=0$ in the control parameter space of the 
scaled densities. The argument of the Bessel function in (\ref{reducedpf}) vanishes in that limit, turning the Bessel function value to unity - the very argument
which is obtained for the absence of any coupling between sites in (\ref{RDNLS}). Therefore the infinite temperature limit corresponds to the case of uncoupled sites,
and the results will be valid for any lattice dimension. Note that the infinite temperature limit $\beta \rightarrow 0$ implies $\beta\mu \rightarrow const$.
With the notation $\mu_l = \mu + \epsilon_l$ and after some simple algebra it follows for infinite temperature
\begin{equation}
\ln \mathcal{Z} = N \ln 2\pi - \sum_{l=1}^N \left( \ln ( \beta \mu_l) + \frac{\beta}{\beta^2 \mu_l^2} \right)
\;.
\label{finalpf}
\end{equation}
With the standard relations $ \mathcal{H}   = ((\frac{\mu}{\beta}\frac{\partial}{\partial \mu} - \frac{\partial}{\partial \beta} ) \ln \mathcal{Z}$ and 
$ \mathcal{A} = -\frac{1}{\beta} \frac{\partial}{\partial \mu } \ln \mathcal{Z}$ we obtain
\begin{equation}
\mathcal{A} = \frac{N}{\beta \mu}\;,\;\mathcal{H} = \frac{N}{\beta^2\mu^2} + \sum_{l=1}^N \frac{\epsilon_l}{\beta \mu_l}\;.
\label{normenergy}
\end{equation}
Note that we used $\mu \sim 1/\beta$ in the infinite temperature limit.

The total norm is not affected by the presence of a potential. The energy is affected, however things are different for the densities.
The second term in the energy expression in (\ref{normenergy}) can be expanded as
\begin{equation}
\sum_{l=1}^N \frac{\epsilon_l}{\beta(\mu+\epsilon_l)} \approx \frac{1}{\beta \mu} \left( \sum_{l=1}^N \epsilon_l - \frac{1}{\mu} \sum_{l=1}^N \epsilon_l^2 \right) \;.
\end{equation}
The first term on the rhs diverges as $\sqrt{N}$ for any finite variance of $\epsilon_l$ - too slow to contribute to the final relation between the densities at
the infinite temperature point (because we have to divide by $N$). The second term on the rhs is proportional to $N/\mu$ - the thermodynamic limit will remain a contribution in the density, however
the infinite temperature limit leads to $1/\mu \rightarrow 0$ and therefore both terms vanish. As a result, for any potential with zero average and finite variance,
the infinite temperature line in the density control parameter space for any lattice dimension is given by
\begin{equation}
y_{\beta=0} = x^2 \;.
\label{betazero}
\end{equation}
Since the energy density is not bounded from above, we conclude that for all densities $y > y_{\beta=0}$ the system will not be described by a Gibbs distribution.

Let us discuss some consequences in the absence of disorder (the presence of disorder will not substantially alter them). 
First, at a given scaled norm density $x$, a homogeneous state $A_l=const$ with constant phases $\phi_l=const$ will correspond
to the lowest energy state which is in the Gibbsean regime. For staggered phases $\phi_{l+1} = \phi_{l}+\pi$ the homogeneous state $A_l=const$ yields a scaled energy density
$y_{st} = 2x+x^2/2$. Therefore we find that for $x \geq 4$ all homogeneous initial states $A_l = const$, regardless of their phase details, are launched in the Gibbsean regime. 
However for $x < 4$ a growing set of homogeneous states with nonconstant phases, in particular the staggered case, are located in the nonGibbsean regime.
We also stress that for all average scaled norm densities $x$ there exist initial states which are inhomogeneous in the amplitudes such that the state will be located
in the nonGibbsean regime. 

If an initial state is in the nonGibbsean regime, we can not conclude much about the nature of its dynamics. It could remain to be strongly chaotic, and described by a negative temperature
Gibbs distribution. It could be also nonergodic, less chaotic, or non-mixing. 

The reported dynamical regimes in the Gibbsean and nonGibbsean are remarkably different \cite{rckgj00,jr04,br07,br08,br09,DB}. While the Gibbsean regime is characterized
by a relatively quick decay into a thermal equilibrium on time scales which are presumably inverse proportional to the largest Lyapunov coefficient, the nonGibbsean regime
is very different. The dynamics is still chaotic, however the system relaxes into a two-component state - condensed hot spots with concentrated norm in them and corresponding high energy,
and cold low energy density regions between them. Some of the results seem to indicate that the system produces as much of a condensate fraction as is needed to keep the remaining noncondensed part
in a Gibbsean regime with infinite temperature$ \beta=0$. There is no evident mixing and relaxation in the condensate fraction.
The condensed hot spots are similar to discrete breathers which are well known to exist in such models \cite{DB}.

Interestingly models without norm conservation also allow for discrete breathers \cite{DB}. With only one conservation law (energy) and one variational parameter (inverse temperature)
the equilibrium state of a Boltzmann distribution is always capable of yielding the prescribed energy density. Still such systems can produce hot spots, or discrete breathers, in thermal
equilibrium in certain control parameter domains \cite{DB}. The remarkable difference to the above cases is, that the condensed hot spots do have a finite life time, and mixing, ergodicity 
and thermal equilibrium are obtained after finite times.

\section{ Selftrapping theorem}
\label{sec4}

The existence of the second conserved quantity $\mathcal{A}$ has also a nontrivial consequence for the decay of  localized initial states or simply wave packets \cite{kkfa08}.
Consider a compact localized initial state with finite norm $\mathcal{A}$ and energy $\mathcal{H}$. Such a state has nonzero amplitudes inside a finite volume only, and strictly
zero amplitudes outside. Note that the theorem can be easily generalized to non-compact localized initial states with properly, e.g.
exponentially, decaying tails.  The theorem addresses the question whether such a state can spread into an infinite volume and dissolve completely into some homogeneous final states.
To measure the inhomogeneity of states we use the participation number (PN)
\begin{equation}
P = \frac{\mathcal{A}^2}{\sum_l A_l^2} \;.
\label{PN}
\end{equation}
This measure is bounded from below $ P \geq 1$ and from above by $P \leq N$ where $N$ is the number of available sites. The lower bound is achieved by concentrating all the available norm onto one single site $A_l = \mathcal{A} \delta_{l,l_0}$. The upper bound
is achieved by a completely homogeneous state $A_l=\mathcal{A}/N$. Note that a typical value of the PN in a thermalized state
is about $N/2$, due to inavoidable fluctuations in the amplitudes on different sites. 

For an infinite system $N \rightarrow \infty$, the PN is unbounded from above. A localized initial state has a finite PN.
If this state evolves and stays localized, its PN stays localized as well. If it manages to spread into the infinitely large reservoir
of the system, and if the densities become on all sites of order $\mathcal{A}/N$, then the PN will be of order $N$. If the PN
stays finite, then a part of the excitation is said to stay localized - either in the area of the initial excitation spot, or in other, possible
migrating, locations. Therefore, the participation number turns to be a useful measure of inhomogeneity of a state, including
localized distributions on zero or also nonzero delocalized backgrounds. It is the more useful as its inverse, up to a constant,
is precisely the anharmonic energy share of the full Hamiltonian (\ref{RDNLS}).

The selftrapping theorem \cite{kkfa08} uses the existence of the second integral of motion - the norm. 
We split the total energy 
$\mathcal{H}= \langle\psi|\mathbf{L}|\psi\rangle + 
H_{NL}$ 
into the sum of its  
quadratic term of order $2$ and its nonlinear terms of order strictly higher than $2$.
Then, $\mathbf{L}$ is a linear operator which is bounded from above and below.
In our specific example, we have  $\langle\psi|\mathbf{L}|\psi\rangle \geq \omega_{m} 
\langle\psi|\psi\rangle= \omega_{m} \mathcal{A}$ 
where $2 + \frac{W}{2} \geq \omega_{m} \geq  -2 -\frac{W}{2}$ and $\omega_m$ is an eigenvalue of $\mathbf{L}$.

If the wavepacket amplitudes spread to zero at infinite time,  
$\lim_{t\rightarrow \infty}( \sup_l  |\psi_l|)=0$. 
Then 
$\lim_{t\rightarrow \infty} ( \sum_l  |\psi_l|^4) <  \lim_{t\rightarrow \infty}
( \sup_l |\psi_l^2|) (\sum_l  |\psi_l|^2)  =0$ since  $\mathcal{A}=\sum_l | \psi_l|^2$ is time invariant.  
Consequently ,  for $t \rightarrow \infty$ we have $\mathcal{H}_{NL}=0$ and $\mathcal{H} 
\leq  \omega_{m} \sum_l |\psi_l|^2=\omega_{m}  \mathcal{A}$. 
Since $\mathcal{H}$ and $\mathcal{A}$ are both time invariant, this inequality should be fulfilled at all times.
However when  the initial amplitude $\sqrt{A}$ of the wavepacket is large enough, it cannot be 
initially fulfilled  because the nonlinear energy diverges as $A^2$  while 
the total norm  diverges as  $A$ only.  
Thus such an initial wavepacket cannot spread to zero amplitudes at infinite time.
This proof is valid for any strength of disorder $W$ including the ordered case $W=0$,
and any lattice. Note that the opposite is not true - i.e., if the wavepacket does not fulfill the criterion for selftrapping
according to the selftrapping theorem, we can only conclude that it may not selftrap in the course of spreading.
We will still coin this regime non-selftrapping.

Let us consider two examples. First, take a single site intial state $\psi_l = \sqrt{A} \delta_{l,0}$. The energy is 
$\mathcal{H}=\epsilon_0 A + \frac{\nu}{2} A^2$, and the norm $\mathcal{A} = A$. A zero amplitude final state has
an upper energy bound of $(W/2+2)A$. For an amplitude $A > A_c$ with $\nu A_c = W+4-\epsilon_0$ the initial
state can not spread into a final one with infinite PN. The PN is bounded from above by $P_{max}$ with
$P_{max}^{-1} = (2\epsilon_0 -W-4)/(\nu A) + 1$.

A second example concerns wave packets excited on many sites.
Assume a wave packet of size $L$ with average scaled energy density $y_0$ and norm density $x_0$. The selftrapping theorem tells
that selftrapping will persist for $y_0 > y_c$ with
\begin{equation}
y_c = \left( \frac{W}{2}+2 \right) x \;.
\label{ST}
\end{equation}

Let us discuss the connection between selftrapping and Gibbs-nonGibbs regimes for the second example of a spreading wave packet
excited initially on many sites. At any time during its spreading (including the initial time) we can trap it with fixed boundaries
at its edges, and address its thermodynamic properties. We also note that if a wave packet spreads, its width $L$ will
increase with time, and the densities $y$ and $x$ will correspondingly drop keeping a linear dependence $y = \frac{y_0}{x_0} x$. 
Then, for $y_0 \leq 0$ the wave packet will stay in the Gibbs regime for all times. Selftrapping does not apply either.
However, for positive $y_0 > 0$ things are more complex. A wave packet will be either all the time nonGibbsean, or enter
a nonGibbsean phase at some later point in time. With Eq. (\ref{ST}) it could also be selftrapped, or not. 

We finally note that a spreading wave packet is representing a nonequilibrium process characterized by corresponding time scales.
The formation of a nonGibbsean density distribution, as reported in \cite{rckgj00,jr04,br07,br08,br09}, takes place on certain
time scales as well. It might therefore well happen that a spreading wave packet does not thermalize quickly enough in its core,
and therefore never enters a nonGibbsean regime despite the fact that it would do so at equilibrium.

\section{Anderson localization}
\label{sec5}
\index{Anderson localization}

For $\nu=0$ with $\psi_{l} = B_{l}
\exp(-i\lambda t)$ Eq.~(\ref{RDNLS-EOM})
is reduced to the linear eigenvalue problem
\begin{equation}
\lambda B_{l} = \epsilon_{l} B_{l} 
- B_{l-1}-B_{l+1}\;.
\label{EVequation}
\end{equation}
All eigenstates are exponentially localized, as first shown by P.W. Anderson \cite{PWA58}.
The normal modes (NM) are characterized by the 
normalized eigenvectors $B_{\bar{\nu},l}$ ($\sum_l B_{\bar{\nu},l}^2=1)$.
The eigenvalues $\lambda_{\bar{\nu}}$ are the frequencies of the NMs.  The width
of the eigenfrequency spectrum $\lambda_{\bar{\nu}}$ of (\ref{EVequation}) is
$\Delta=W+4$ with $\lambda_{\bar{\nu}} \in \left[ -2 -\frac{W}{2}, 2 + \frac{W}{2}
\right] $. While the usual ordering principle of NMs is with their increasing eigenvalues, here we adopt
a spatial ordering with increasing value of the center-of-norm coordinate $X_{\bar{\nu}}= \sum_l l B^2_{\bar{\nu},l}$.

The asymptotic spatial decay of an eigenvector is given by $B_{\bar{\nu},l} \sim
{\rm e}^{-|l|/\xi(\lambda_{\bar{\nu}})}$ where 
$\xi(\lambda_{\bar{\nu}})$ is the localization length \index{localization length} and
$\xi(\lambda_{\bar{\nu}}) \approx
24(4-\lambda_{\bar{\nu}}^2)/W^2$ for weak disorder $W \leq 4$ \cite{KRAMER}. 

The volume occupied by a given eigenstate is denoted by $V_{\bar{\nu}} \sim \xi_{\bar{nu}}$ (for details see \cite{tvlmvisf14}).
The average spacing $d$ of eigenvalues of neighboring NMs
within the range of a localization volume is of the order of $d \approx \Delta / V$,
which becomes $d \approx \Delta W^2 /300 $ for weak disorder.
The two scales $ d \leq \Delta $ are expected to determine the
packet evolution details in the presence of nonlinearity.

Due to the localized character of the NMs, any localized wave packet with size $L$ which is launched into
the system for $\nu=0$ , will stay localized for all times. 
We remind that Anderson localization is relying on the phase coherence of waves. Wave packets which are trapped due
to Anderson localization correspond to trajectories in phase space evolving on tori, i.e. they evolve quasi-periodically in time.

Finally, the linear wave equations constitute an integrable system with conserved actions where the dynamics happens to be on quasiperiodic tori in phase space.
This can be safely stated for any finite, whatever large, system.

\section{Disorder +  Nonlinearity }
\label{sec6}
\index{Kolmogorov-Arnold-Moser (KAM) theorem}
\index{nonlinearity}

In the presence of nonlinearity
the equations of motion of (\ref{RDNLS-EOM})  in normal mode space read
\begin{equation}
i \dot{\phi}_{\bar{\nu}} = \lambda_{\bar{\nu}} \phi_{\bar{\nu}} + \nu \sum_{\bar{\nu}_1,\bar{\nu}_2,\bar{\nu}_3}
I_{\bar{\nu},\bar{\nu}_1,\bar{\nu}_2,\bar{\nu}_3} \phi^*_{\bar{\nu}_1} \phi_{\bar{\nu}_2} \phi_{\bar{\nu}_3}\;
\label{NMeq}
\end{equation}
with the overlap integral 
\begin{equation}
I_{\bar{\nu},\bar{\nu}_1,\bar{\nu}_2,\bar{\nu}_3} = 
\sum_{l} B_{\bar{\nu},l} B_{\bar{\nu}_1,l} 
B_{\bar{\nu}_2,l} B_{\bar{\nu}_3,l}\;.
\label{OVERLAP}
\end{equation}
The variables $\phi_{\bar{\nu}}$ determine the complex time-dependent amplitudes of
the NMs.

The frequency shift of a single site oscillator induced by the nonlinearity is
$\delta_l = \nu |\psi_l|^{2}  \approx x$. 
As it follows from (\ref{NMeq}), nonlinearity induces an interaction between NMs.
Since all NMs are exponentially localized in space, each normal mode is effectively coupled to a finite
number of neighboring NMs, i.e. the interaction range is finite. However the strength of the coupling
is proportional to the norm density $n = |\phi|^2$. Let us assume that a wave packet spreads.
In the course of spreading its norm density will become smaller. Therefore the effective coupling strength
between NMs decreases as well. At the same time the number of excited NMs grows.
One possible outcome would be: (I) that after some time the coupling will be weak enough
to be neglected. If neglected, the nonlinear terms are removed, the problem is reduced to an integrable linear wave equation,
and we obtain again Anderson localization. That implies that the trajectory happens to be on a quasiperiodic torus -
on which it must have been in fact from the beginning. It also implies that the actions of the linear wave equations are not strongly varying in the nonlinear case,
and we are observing a kind of anderson localization in action subspace. 
Another possibility is: (II) that spreading continues for all times. That would imply that the trajectory does not
evolve on a quasiperiodic torus, but instead evolves in some chaotic part of phase space.
This second possibility (II) can be subdivided further, e.g. assuming that the wave packet will exit, or enter, a Kolmogorov-Arnold-Moser (KAM) regime of mixed phase space, or stay all the time outside such a perturbative KAM regime. In particular if the wave packet dynamics will
enter a KAM regime for large times, one might speculate 
that the trajectory will
get trapped between denser and denser torus structures in phase space after some spreading, leading again
to localization as an asymptotic outcome, or at least to some very strong slowing down of the spreading process. 

Published numerical data \cite{PS08,fks08,skkf09,mjgksa10,blskf11,tvljdbdokcssf10}  (we refer to \cite{tvlmvisf14,sf14} for more
original references
and details of the theory)
show that finite size initial wave packets i) stay localized if $ x \ll d$ and display regular-like (i.e. quasiperiodic as in the KAM regime ) dynamics, which is numerically hardly distinguishable from very slow chaotic dynamics with subsequent spreading on unaccessible time scales); ii) spread subdiffusively
if $ x\sim d$
with the second moment of the wave packet $m_2 \sim t^{\alpha}$  and chaotic dynamics inside the wave packet core;
segregate into a two component field with a selftrapped component, and a subdiffusively spreading part for $ x > \Delta$.

Spreading wave packets reduce their densities $x,y$ in the course of time, and may reach regime i), without much change in their
spreading dynamics. Therefore we can conclude that regime i) is at best a KAM regime, with a finite probability to launch a wave packet on
a KAM torus, and a complementary one to observe spreading \cite{mvitvlsf11}. Anderson localization is restored in that probabilistic
way, as the probability to stay on a KAM torus will reach value one in the limit of vanishing nonlinearity.
Spreading wave packets, when launched in a domain of positive energy densities $y$, are either from the beginning in the 
nonGibbsean regime, or have to reach it at some later point in time. Assuming that the wave packet has enough time
to develop nonGibbsean structures, one should observe selftrapping. Published numerical studies did not focus on this
issue, in particular for parameter values which do not satisfy the seltrapping theorem.
The reported numerical selftrapping dynamics is in full accord with the results of the selftrapping theorem. 

The most interesting result concerns the spreading dynamics and the exponent $\alpha$. The assumption of strong chaos - i.e.
the dephasing of normal modes on times scales much shorter than the spreading time scales - leads to $\alpha=1/2$ \cite{sf10,tvljdbdokcssf10,dmb14}
This exponent can be numerically observed, but only as an intermediate (although potentially extremely long lasting) regime
of strong chaos:
\begin{equation}
\alpha=\frac{1}{2}\;,\; x > d \;:\; {\rm strong \; chaos } \;.
\label{sc}
\end{equation}
In the asymptotic regime of small densities, instead the regime of weak chaos is observed:
\begin{equation}
\alpha=\frac{1}{3}\; , \;  x < d  \;:\; {\rm weak \; chaos } \;.
\label{wc}
\end{equation}
Perturbation theories show that in that regime not all normal modes are resonant and chaotic, but only a fraction of them \cite{fks08,skkf09,sf10,sf14}. Correcting the theory of strong chaos with the probability of resonances $\mathcal{P}_r$, the asymptotic
value $\alpha = 1/3$ is obtained \cite{sf10,tvljdbdokcssf10}, and a summary of the results reads as follows \cite{sf10,sf14}:
\begin{equation}
D \sim (\mathcal{P}_r(x)  x \langle I \rangle )^2\;,\; \mathcal{P}_r = 1-{\rm e}^{-C x}\;,\;
C \sim \frac{ \xi^2 \langle I \rangle}{ d }\;.
\label{ent5}
\end{equation} 
The corresponding nonlinear diffusion equation \cite{NLD} for the norm density distribution
(replacing the lattice by a continuum for simplicity, see also \cite{ark10}) uses the diffusion coefficient $D(x)$ :
\begin{equation}
\partial_t x = \partial_{\bar{\nu}}(D \partial_{\bar{\nu}} x) \;.
\label{ent4}
\end{equation}
In the regime of strong
chaos it follows $ D \sim x^2$, and in the regime of weak chaos - $D \sim x^4$. 
It is straightforward to show that the exponent $\alpha$ satisfies the relations (\ref{sc}) and (\ref{wc}) respectively \cite{sf10,Mul10,tvljdbsf13,sf14}.
We can also conclude
that a large system at equilibrium will show a conductivity which is proportional to $D$.

\section{A discussion in the light of many body localization}

Let us discuss the expected dynamical regimes of an infinitely large lattice excited to some finite densities. In the nonGibbsean regime $y > x^2$ the dynamics is
known to be nonergodic up extremely long time scales. At the same time the Gibbsean phase is characterized by two different regimes - strong chaos at large densities, and
weak chaos at small densities. Strong chaos implies that all normal modes are resonant and lead to chaos and mixing. Therefore we could assume that strong chaos is ergodic.
For small enough density $x$ we enter the weak chaos regime, where not all normal modes are resonant and lead to chaos and mixing at the same time. Therefore it could be
possible that this regime is nonergodic, despite the fact that it characterized by a finite conductivity.

Now we recall the main statements from many body localization. This theory deals with a quantum system of many interacting particles in the presence of disorder.
It was initially developed for fermions \cite{dmbilabla06}. The system is considered in a spatially continuous system. At a given particle density, the conductivity is predicted to be zero
up to a nonzero critical energy density. Above this transition point, the system exhibits many body states which conduct in a nonergodic (multifractal) fashion. At even larger
energy densities the system finally exhibits ergodic metallic states. Later this theory was also developed for bosons, which is the quantum counterpart of our classical model \cite{bosonmbl}.
In that case, again the system is characterized by a finite energy density (at fixed particle density) below which all states are many body localized. Above that critical density states are
again extended but nonergodic and fractal. 

The cricital many body localization energy density scales to zero in the classical limit. This is consistent with the fact that a wave packet spreads to infinity (assuming that it indeed does so),
as the wave packet is characterized by finite densities at any finite time, and approaches zero density in the limit of infinite time. It is therefore tempting to associate the regime of weak chaos with
the nonergodic but metallic regime of the quantum theory. The test of this possibility is therefore one of the challenging future tasks to be explored.


\begin{thebibliography}{99.}%

\bibitem{PWA58} P. W. Anderson, Phys. Rev. {\bf 109} 1492 (1958).


\bibitem{Exp} T. Schwartz, G. Bartal, S. Fishman and M. Segev, Nature
\textbf{446} 52 (2007) 

\bibitem{Exp2}Y. Lahini, A. Avidan, F. Pozzi, M. Sorel,
R. Morandotti, D. N. Christodoulides and Y. Silberberg,
Phys. Rev. Lett. \textbf{100} 013906 (2008).

\bibitem{BECEXP} D. Clement A. F. Varon, J. A. Retter, L. Sanchez-Palencia,
A. Aspect and P. Bouyer, New J. Phys. \textbf{8} 165 (2006);
L. Sanches-Palencia D. Clement, P. Lugan, P. Bouyer, G. V. Shlyapnikov and
A. Aspect, Phys. Rev. Lett. \textbf{98} 210401 (2007); J. Billy, V. Josse,
Z. Zuo, A. Bernard, B. Hambrecht, P. Lugan, D. Clement, L. Sanchez-Palencia,
P. Bouyer and A. Aspect, Nature {\bf 453}, 891 (2008); G. Roati, C. D'Errico,
L. Fallani, M. Fattori, C. Fort, M. Zaccanti, G. Modugno, M. Modugno and
M. Inguscio, Nature {\bf 453}, 895 (2008).


\bibitem{oberthaler}
O. Morsch and M. Oberthaler, Rep. Prog. Phys. \textbf{78} 176 (2006).

\bibitem{yskgpa03} Yu. S. Kivshar and G. P. Agrawal, {\sl Optical Solitons:
From Fibers to Photonic Crystals} (Academic Press, Amsterdam, 2003).

\bibitem{rckgj00}
K. \O. Rasmussen, T. Cretegny, P. G. Kevrekidis and N. Gr{\o}nbech-Jensen, Phys. Rev. Lett. {\bf 84} 3740 (2000).

\bibitem{jr04}
M. Johansson and K. \O. Rasmussen, Phys. Rev. E {\bf 70} 066610 (2004).

\bibitem{br07}
B. Rumpf, EPL {\bf 78} 26001 (2007).

\bibitem{br08}
B. Rumpf, Phys. Rev. E {\bf 77} 036606 (2008).

\bibitem{br09}
B. Rumpf, Physica D {\bf 238} 2067 (2009).

\bibitem{DB} S. Flach and C. R. Willis, Phys. Rep. \textbf{295} 181 (1998);
D. K. Campbell, S. Flach and Yu. S. Kivshar, Phys. Today {\bf 57} (1), 43 (2004);
S. Flach and A. V. Gorbach, Phys. Rep. \textbf{467}, 1 (2008).

\bibitem{dmb14}
D. M. Basko, Phys. Rev. E {\bf 89} 022921 (2014).

\bibitem{kkfa08} G. Kopidakis, S. Komineas, S. Flach and S. Aubry,
Phys. Rev. Lett. \textbf{100} 084103 (2008). 

\bibitem{KRAMER}
B. Kramer and A. MacKinnon, Rep. Prog. Phys. \textbf{56} 1469 (1993).

\bibitem{tvlmvisf14}
T. V. Laptyeva, M. V. Ivanchenko and S Flach, J. Phys. A {\bf 47}, 493001 (2014).

\bibitem{PS08} A. S. Pikovsky and D. L. Shepelyansky,
Phys. Rev. Lett. \textbf{100} 094101 (2008).

\bibitem{fks08} S. Flach, D. Krimer and Ch. Skokos,
Phys. Rev. Lett. \textbf{102} 024101 (2009).

\bibitem{skkf09} Ch. Skokos, D. O. Krimer, S. Komineas and S. Flach,
Phys. Rev. E \textbf{79} 056211 (2009).

\bibitem{mjgksa10}
M. Johansson, G. Kopidakis and S. Aubry,
Europhys. Lett. {\bf 91} 50001 (2010).

\bibitem{blskf11}
J. Bodyfelt, T. V. Laptyeva, Ch. Skokos. D. Krimer and S. Flach, Phys. Rev. E {\bf 84}, 016205 (2011).

\bibitem{tvljdbdokcssf10}
T. V. Laptyeva, J. D. Bodyfelt, D. O. Krimer, Ch. Skokos and S. Flach,
EPL {\bf 91}, 30001 (2010).

\bibitem{sf14}
S. Flach, arxiv1405.1122 .

\bibitem{mvitvlsf11}
M. V. Ivanchenko. T. V. Laptyeva and S. Flach,
Phys. Rev. Lett. {\bf 107} 240602 (2011).

\bibitem{sf10}
S. Flach, Chem. Phys. {\bf 375}, 548 (2010).

\bibitem{NLD} 
Y.B. Zel’dovich and Y.P. Raizer, {\sl Physics of Shock Waves and High-Temperature Hydrodynamic Phenomena}, Academic Press, New York, 1966;
Y.B. Zel’dovich and  A. Kompaneets, {\sl Collected Papers of the 70th Anniversary of the Birth of Academician A. F. Ioffe}, Moscow, 1950;
G.I. Barenblatt, Prikl. Mat. Mekh. {\bf 16}, 67 (1952).

\bibitem{ark10}
A.R. Kolovsky, E.A. Gomez and H.J. Korsch, Phys. Rev. A {\bf 81}, 025603 (2010).

\bibitem{Mul10} M. Mulansky and A. Pikovsky, EPL {\bf 90}, 10015 (2010).

\bibitem{tvljdbsf13}
T. V. Laptyeva, J. D. Bodyfelt and S. Flach,
Physica D {\bf 256-257}, 1 (2013).

\bibitem{dmbilabla06}
D. M. Basko, I. L. Aleiner and B. L. Altshuler,
Annals of Physics {\bf 321} 1126 (2006). 

\bibitem{bosonmbl}
I. L. Aleiner, B. L. Altshuler and G. V. Shlyapnikov, Nature Physics {\bf 6}, 900 (2010).



\end{thebibliography}
\end{document}